# Adaptive Multi-Agent E-Learning Recommender Systems

Nethra Viswanathan
*Department of Computing*
Bournemouth University
Poole, United Kingdom
s5227228@bournemouth.ac.uk

*Abstract*— Educational recommender systems have become a necessity in the recent years due to overload of available educational resource which makes it difficult for an individual to manually hunt for the required resource on the internet. E-learning recommender systems simplify the tedious task of gathering the right web pages and web documents from the scattered world wide web repositories according to every users' requirements thus increasing the demand and hence the curiosity to study them. Retrieval of a handful of recommendations from a very huge collection of web pages using different recommendation techniques becomes a productive and time efficient process when the system functions with a set of cooperative agents. The system is also required to keep up with the changing user interests and web resources in the dynamic web environment, and hence adaptivity is an important factor in determining the efficiency of recommender systems. The paper provides an overview of such adaptive multi-agent e-learning recommender systems and the concepts employed to implement them. It precisely provides all the information required by a researcher who wants to study the state-of-the-art work on such systems thus enabling him to decide on the implementation concepts for his own system.

## I. Introduction

The paper provides a detailed review on the different implementation techniques employed in the state-of-the-art adaptive multi-agent E-learning recommender systems along with their benefits and drawbacks . E-learning recommender systems are seen on websites recommending educational resources from learning object repositories (LOR) and training materials in the world wide web fraternity to students and researchers based on several factors. The factors include interests and preferences, their learning level, their purchase/browsing history in the system, time spent with each resource and also based on the interests of other users who have similar learning objectives as theirs [1]. Educational domain is chosen because there is a great number of diverse resources that can contribute to teaching-learning process. This is where users' preferences and restrictions need to be considered and hence can be modelled at different levels of complexity showing scope for intelligence [2]. Here agents are considered to be autonomous entities such as software programs or robots [3]. Some of the approaches followed for recommendation of learning objects are content based approach, collaborative approach and hybrid approach which are explained later in the paper.

Can a multi-agent web recommender system with multiple recommendation algorithms practically perform better than a single agent recommender system based on a single algorithm [1]? Neto [1] had proposed to prove that Multi-Agent systems are superior than single agent systems in recommending the appropriate resources with a negligible turnaround time. Adding to this, Morais et al. [4] have proved by evaluating two datasets that Multi-Agent system outperforms individual algorithm execution using Associative Rules (AR) and collaborative filtering (CF) algorithms for recommendation. We also understand that a single agent implementing one of the said recommendation techniques performs a more restricted search for resources than a multi-agent which employs multiple techniques to provide simultaneous recommendations thus providing best fit results [5].

The rest of the paper is organised as follows: The next section gives an overview on the adaptive multi-agent systems and their advantages as experienced by researchers. Section 3 briefly explains the available agent architectures and their application in the cited papers. Section 4 explains and compares the different recommendation techniques implemented to fetch the best results customised for every user who accesses the e-learning system. Section 5 provides a brief explanation on the existing research work on the different concepts employed in implementing adaptive multi-agent e-learning recommender systems and their advantages and drawbacks. We end the paper with a conclusion and future work section to provide a summary of the obtained knowledge from cited papers and areas for future study on the subject.

## II. Adaptive Multi-Agent Systems

Multi-agent systems possess the advantage of providing simultaneous recommendations with cooperation by distributing the tasks among them to clearly identify problems that must be resolved by each agent [4], [6]. Students with different capabilities and interests use e-learning resources and their interests keep changing with time, hence self-adjusting or adaptive recommendations for every user is fundamental for covering the updated interests of every student in every level of education. Adaptive multi agent recommender systems operate with a set of agents interacting in a collaborative and cooperative manner and executing incremental algorithms to make bids and provide next set of best-fit recommendations to the user [5]. Fig. 1 explains the main components for the implementation of adaptivity in one of the intelligent agents of a recommender system where the system keeps updating the trained model with the new incoming ratings for resources. The agent's recommendation algorithm uses the new ratings of resources to retrain the recommender model and obtain new prediction values for the unrated items thus keeping up with the changing environment [7].

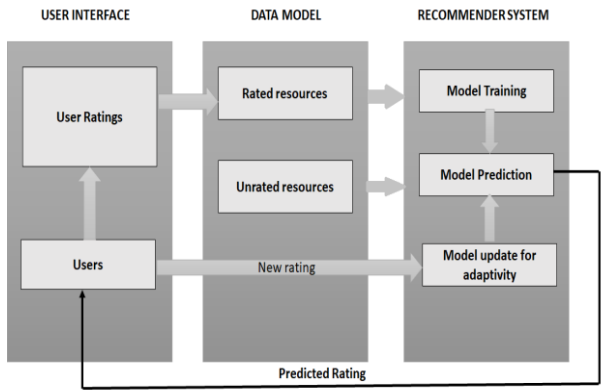

Fig. 1. Main components of Adaptive recommendation agent

There are several agent architectures which facilitate the execution of such systems in a well organised manner which is described in the next section..

## III. AGENT ARCHITECTURES

The agent architecture plays an important role in determining the implementation plan of the agent in solving the problem. It acts as a canvas on top of which the different functionalities or modules are implemented in layers. The overall classification of agent architectures, as per Siebers and Aickelin [8], is as follows:

### A. Logic based agent Architecture

Agents are specified in a rule-like language and decision making is implemented through logical deduction.

### B. Reactive Agent Architecture

Agents aim to attain goals which are decomposed into layers of inter-dependant smaller modules. The architecture is used when agents are required to mimic very complex adaptive environments [9].

### C. Layered Architecture:

Agents operate with separate layers for reactive, pro-active and social behaviour which can be connected horizontally or vertically with respect to the input/output [10].

### D. Belief-Desire-Intention architecture:

Beliefs represent the information sources which are prone to changes, Desires represent the objectives of the agents and Intentions represent the chosen goals. The architecture suits the development of any adaptive system where goals are not consistent and hence the changing desires express the goals better [11].

Casali et al. [2] have proposed an adaptive multi-agent learning object retrieval mechanism using Belief-Desire-Intention (BDI) architecture for one of its agents, Personalised Search (PS) Agent, for providing intelligent recommendations using recommendation techniques explained in the forthcoming section. The authors have designed six agents, namely User Interface agent, Semantic refiner agent, User profile agent, Search agents which receives input from the above agent and hands over to the PS Agent. The metadata of the learning objects constitutes the Beliefs and users' learning objectives constitute the Desires. The mapping between the two is calculated in order to rank the recommended learning objects as per preference. The drawback in the paper is that many of the learning objects were not accessible and lack of metadata of learning objects which meant generating recommendations from only a subset of learning objects

## IV. RECOMMENDATON TECHNIQUES

Recommendation techniques form the brain of any recommender system as it determines the efficiency factor of the developed system. Collaborative filtering, content-based filtering and hybrid filtering algorithms are the most widely used recommendation techniques among the cited papers in this research. The below furnished table TABLE I provides the definitions and vivid differences between the different recommendation techniques used by state-of-the-art educational recommender systems [5].

TABLE I. COMPARISON OF RECOMMENDATION TECHNIQUES

| Content-based Filtering | Collaborative Filtering |
|---|---|
| The technique provides recommendations by matching the user preferences for resource features thus fetching the best recommendations for the user. The user gets recommendations based on his previous feedback. | The technique predicts user preferences as a linear, weighted combination of like-minded users' preferences and thus provides the best recommendations for the user. |
| The benefit of the technique is the fact that recommendations are provided based on the user's previous product ratings and hence guaranteed to fetch the customised match for the user. | The benefit of the technique is that a user is presented with recommendations which he might not have expected but is of great interest because of one of his favourite factors in the resource which was rated high by another like-minded user. |
| The disadvantage is that the system needs constant feedback from users in order to provide update recommendations as per the changing interests of the user which creates overhead for the system. | The disadvantage is that the system needs a huge dataset of active user ratings in order to fetch similar users' ratings for providing recommendations. |

The hybrid recommendation technique provides the benefits of both of the former techniques and hence considered a good option for an efficient recommender system. However, it also has the disadvantages of both of the former systems as it fetches recommendations following both the techniques. Additionally speed of computation is expected to be low because of implementation of two techniques in place of one unless it is handled by the algorithm [5].

Selmi et al. [5] discuss about systems which employ collaborative and content-based filtering in parallel thus reducing the turnaround time of the recommender which produces good quality results in less execution time. Rodriguez Marin and Neto [1], [6] support content-based filtering mechanism for generating recommendations. The system employs nine agents namely, Clustering agent, Content based recommender, Collaboration based recommender, knowledge-based recommender, hybridisation agent, ontological agent which infers knowledge from ontology, communication agents between user and LO , agent for web service bridging between MAS and web application. They also back knowledge based and hybrid recommendation techniques which ensure recommendations involving all the possible search criteria. Knowledge based search involves the similarity of search topic with navigation history and previous elections. The quality of recommendations is assessed by precision and recall formulae.

## V. Related Work And Discussion

Multiple papers have been published on adaptive multi-agent educational recommenders in the past decade. According to the study [12], the challenge with adaptive MAS is to keep track of the change of behaviour of agents with respect to the context and change of resources over time in order to keep providing relevant recommendations at all times. Neto [1] also argues that in an open, heterogenous and dynamic multi agent environment, it is difficult to maintain trust among the interacting agents which plays an important role in bringing out an efficient output. Hence modelling and evaluating trust relationship between the agents is a major challenge. On the other hand, Bednarik et al. [13] confronts that one of the important functionalities of multi agent education systems is automatic collection and updating of information on websites based on the current educational trend which is very useful for recommendations. They also describe that multiple agents provide the means to manage complexity and uncertainty of the domain in the context of education system architecture. Thus, we see that the beneficial aspects of adaptive MAS for educational resource recommendation can be employed by balancing the mentioned drawbacks of the systems .

Regarding importance of personalisation and adaptivity of a recommendation system, Li et al. [14] have proved through a survey that over 93% of the students who received E-learning recommendations were fully satisfied with the personalised recommendations provided to them and the satisfaction rate is higher than that received for "One size fits all" or non-adaptive recommender systems which did not account for changes in preferences of users. Yang et al. [15] propose an adaptive e-learning system where the user profile keeps updating with changes in user preferences and behaviour with time thus maintaining the adaptivity factor to provide the best fit recommendation at all times. We see that the authors recommend an adaptive system for a better user experience and satisfaction.

According to Melesko and Bednarik [13], [16], Felder-Silvermann Learning style model serves the best for providing E-learning recommendations especially for engineering courses. The learning styles in the said model are classified into four learning schemes, namely sensing, visuals, active and sequential. Each learning object (LO) in the learning object repository (LOR) has the learning scheme included in its metadata and the preferred learning scheme is obtained from the user through his profile information recorded as part of E-learning registration form, preferably Soloman-Felder questionnaire. The study [16] also employs five agents, one of which is a Learning style identification agent in order to identify and map users' learning style with learning objects' metadata thus generating top recommendations based on learning style followed by search topic and user's interests collected through various means.

Yang and Wu [15] propose an attribute-based search mechanism to find adaptive learning objects using Kolb's learning style model for learning scheme classification and ant colony optimisation is used to recommend adaptive LO using web portal for learners. Kolb's learning style model is classified into four learning schemes: Diverging, converging, Assimilating and Accommodating which are assigned to LOs accordingly. Both of the said learning style models are widely used for generating recommendations based on the respective learning schemes.

Optimisation techniques, say Ant Colony optimisation and Genetic Algorithms (GA) , Particle Swarm Optimisation(PSO) have been used by Yang et al. [15] and Li et al. [14] respectively. According to study [15], most personalised learning mechanisms neglect the relationship between learners attributes and attributes of learning objects, however ant colony algorithm is able to accomplish the same for the proposed system. Ant colony uses metaheuristics to cooperatively find high quality solutions to complex combinatorial optimization problems within acceptable time. Ants lay pheromone and heuristic information to mark trails and thus finds shortest path to food. [14] propose GA and PSO algorithms to get close to optimal recommendations which meet individual user requirements. To make the system adaptive, the feedback of the user after accessing each learning material is passed back to the system to self-adjust the user profile based on changing interests thus facilitating adaptive recommendations. When there are less than 300 learning objects , the results indicate that Particle Swarm Optimisation (PSO) is more effective than Genetic Algorithm (GA). However the paper fails to provide an optimal solution for recommendations in terms of computational time irrespective of number of available resources.

RODRÍGUEZ MARÍN et al. [6] discuss about an agent performing clustering technique K-means used to group similar LOs based on ranking and thus deliver apt resources for a specific student with good precision. Clustering is performed on LO metadata while a new LO is added to LOR and the search topic is retrieved from the found cluster. Here clustering reduces the search cost of the recommendation algorithms since the search area is narrowed down to selected clusters in place of the whole database.

## VI. Conclusion And Future Work

We have presented a brief explanation on adaptive multi-agent systems with a pictorial depiction of some of the main components of the system. We have then discussed on the concepts employed in the state-of-the-art adaptive multi-agent e-learning recommender systems and their benefits and drawbacks. We have also described the available agent architectures which act as a canvas for development of the system. We have finally described the commonly used recommendation techniques with a comparison table for a good understanding. As stated in discussion, we see some gaps which need improvement in the cited papers where the computational time is compromised with increase in the number of available resources. When the drawback is overcome, we see lack of feedback from agents which reduces coordination in the system. Adding to this, many learning objects in repositories are not accessible or lacks metadata which is a major drawback for provision of good recommendations. On the other hand, we have also seen useful concepts like clustering and optimisation techniques which contribute in reducing computation time without compromising on the quality of results. In order to reduce the gaps we plan to develop an adaptive MAS for the E-learning platform with clustering and optimisation techniques using intelligent agents. We also plan to improve the communication among the agents by introducing a trust factor for efficient operation. A successfully designed system is bound to result in an adaptive and efficient web resource recommender for all kinds of students at different levels of education as well as for researchers.